\documentclass[12pt]{iopart}
\usepackage{epsfig}
\usepackage{graphicx}
\def\n{\bar n}

\begin{document}

\title{Non-order 
parameter Langevin equation for a bounded Kardar-Parisi-Zhang universality 
class}
%\article[Non-order parameter Langevin]{Paper}{Non-order 
%parameter Langevin equation for a bounded Kardar-Parisi-Zhang universality 
%class}
%\draftcopyName{\today}{130}
%\draftcopySetScale{80}
%\draftcopyFirstPage{1}
%\draftcopyLastPage{2}
%\draftcopySetGrey{0.95}

\author{Omar Al Hammal, Francisco de los Santos, and Miguel A. Mu\~ noz} 
\address{Instituto de F{\'\i}sica Te\'orica y Computacional Carlos I
 and \\Departamento de Electromagnetismo y F{\'\i}sica de la Materia,
 \\ Universidad de Granada, Fuentenueva s/n, 18071 Granada, Spain}
\eads{omar@onsager.ugr.es, dlsantos@onsager.ugr.es and mamunoz@onsager.ugr.es}

\begin{abstract}

 We introduce a Langevin equation describing the pinning-depinning
 phase transition experienced by Kardar-Parisi-Zhang interfaces in the
 presence of a bounding ``lower-wall''. This provides a continuous
 description for this universality class, complementary to the
 different and already well documented one for the case of an
 ``upper-wall''. The Langevin equation is written in terms of a field
 that is not an order-parameter, in contrast to standard approaches,
 and is studied both by employing a systematic mean-field
 approximation and by means of a recently introduced efficient
 integration scheme.  Our findings are in good agreement with known
 results from microscopic models in this class, while the numerical
 precision is improved. This Langevin equation constitutes a sound
 starting point for further analytical calculations, beyond
 mean-field, needed to shed more light on this poorly understood
 universality class.
\end{abstract}
\submitto{J. Stat. Mech.}
\maketitle

\date{today}
\tableofcontents
\section{Introduction}

It was shown a few years ago that the introduction of a limiting or
bounding wall into a Kardar-Parisi-Zhang (KPZ) interface model
\cite{KPZ,HZ} leads to quite different phenomenologies depending on
whether the wall is an ``upper'' or a ``lower'' one
\cite{Review,MN1,MN2,MN3}. This result, the origin of which can be
traced back to the absence of height-inversion symmetry in KPZ
interfacial dynamics \cite{Review,MN3}, has been verified for several
discrete interfacial KPZ-like models \cite{Review,MN3,Fattah}.  In any
of these cases, once a limiting wall is introduced, there are two
different phases: a {\it depinned} one in which the KPZ-interface
moves freely away from the wall, and a {\it pinned} one, with a finite
stationary average distance from the wall. Separating these two phases
there is a nonequilibrium phase transition, whose criticality
encompasses also that of synchronizing extended systems
\cite{Ahlers,Romu}, nonequilibrium wetting phenomena
\cite{Lisboa,Haye1,FSS-MN}, transitions occurring in DNA
alignment problems \cite{Review,MN3}, phenomena related to Burgers'
turbulence \cite{Review}, bounded directed-polymers in random-media
\cite{Review}, etc.  Characterizing and distinguishing between the two
above-mentioned pinning-depinning universality classes, with an upper
or a lower wall respectively, is therefore a relevant task in many
different contexts, as well as a chief theoretical problem.

In terms of Langevin equations, an ideal framework to discuss
universality issues, KPZ-like interfaces are described by the
celebrated and profusely studied KPZ equation
\cite{KPZ}:
\begin{equation}
{\partial{h(r,t)} \over {\partial{t}}}=a+\lambda(\nabla
h(r,t))^{2}+ D \nabla^{2} h(r,t)+\sigma \eta(r,t),
\label{kpz}
\end{equation}
where $h(r,t)$ is the interface height at position $r$ and time $t$,
$\lambda$, D, and $\sigma$ are constants, and $\eta$ is a Gaussian white
noise. In what follows, and without loss of generality, we take
$\lambda= + D$. Alternatively, we could also have fixed a given type
of limiting wall, for instance, a lower, rigid substrate on top of
which the interface grows, and observe the two different classes of
depinning transitions under scrutiny depending on the sign of
$\lambda$ \cite{nota}.  

Let us consider now equation (\ref{kpz}) in the presence of an
exponential, upper wall, i.e. including the additional term $-b \exp(q
\ h)$, with $b>0$ and $q>0$.  A transition takes place from a regime
characterized by depinned interfaces, flowing to minus infinity for
sufficiently negative values of $a$, to one with interfaces pinned to
the wall, with exponentially cutoff positive values of $h$, above a
certain threshold $a_c$.  This can also be visualized by performing a
Cole-Hopf transformation, $n=\exp(h)$ \cite{nota2}, which leads to,
\begin{equation}
{\partial{n(r,t)} \over {\partial{t}}}=D \nabla^{2} n(r,t) +a n(r,t)- b
n(r,t)^{1+q}+ \sigma n(r,t) \eta(r,t).
\label{mn1}
\end{equation}
This is a {\it multiplicative noise} equation (interpreted in the
Stratonovich sense \cite{Ito,VK}) defining a, well established by now,
phase transition with a corresponding set of critical exponents that
characterizes the so-called multiplicative noise 1 (MN1) universality
class \cite{MN1,MN2,MN3,MN4}. At the transition point and in the
depinned phase the stationary average value of the order-parameter,
$\bar{n}$, is zero, while it is non-vanishing above the transition
point. The critical exponents and other universal features in this
class are not affected by the value of $q$, i.e. by the
``impenetrability degree'' of the wall; indeed, in microscopic models
the wall is typically a hard substrate, corresponding to $q\rightarrow
\infty$.

On the other hand, by introducing in equation (\ref{kpz}) a 
lower wall, $b \exp(- q h)$, hindering the interface height to take
negative values, depinned interfaces flow towards plus infinity. In
this case, a natural order-parameter, equivalent to that for the
upper-wall class, going to zero at the transition point, is
$m=\exp(-h)$, and the corresponding depinning transition is in the
so-called multiplicative noise 2 (MN2) universality class
\cite{MN3,Review,Fattah,Haye-MF,FSS-MN}. 

Consequently, we perform the change of
variables $m=\exp(-h)$ to obtain \cite{Fattah,factor2}:
\begin{equation} {\partial{m(r,t)} \over {\partial{t}}}=D\nabla^{2}
m -2D {(\nabla m)^2 \over m} - a m - b m^{1-q} + \sigma m \eta(r,t),
\label{mn2bis}
\end{equation}
where some space and time dependencies have been omitted to simplify
the notation. Observe that owing to the $(\nabla m)^2 /m$ term the
equation becomes singular above the transition point, where $\bar
m=0$.  Equation (\ref{mn2bis}) was studied in detail, both numerically
and using mean-field approaches, in \cite{Fattah}, but no sound result
could be obtained owing to the presence of the singular
gradient term.  Therefore, all the known results for this
universality class come from numerical \cite{Fattah,FSS-MN} as well as
some analytical (mean-field like) \cite{Haye-MF} studies of discrete
microscopic models. Let us also stress that direct numerical
integrations of KPZ-like Langevin equations (before applying the
Cole-Hopf transformation) are uncontrollable due to well-documented
numerical instabilities \cite{KPZproblem}.

Aimed at filling the gap between discrete and continuous levels of
description for the MN2 universality class, in this article we show
that the MN2 phenomenology can be captured by an alternative,
multiplicative-noise, Langevin equation.  To that purpose, we take the
KPZ equation in the presence of a lower-wall and perform the
change of variables, $n=\exp(h)$, leading to
\begin{equation}
{\partial{n(r,t)} \over {\partial{t}}}=D\nabla^{2} n +a n+ b n^{1-q}
+ \sigma n \eta(r,t),
\label{mn2}
\end{equation}
which takes a particularly simple form for $q=1$ although, as in the
MN1 class, the precise value of $q>0$ is not expected to affect
universal features. Note the remarkable difference between equations
(\ref{mn1}) and (\ref{mn2}): while in the former $n$ is an
order-parameter for the MN1 transition, in equation (\ref{mn2}) for
the MN2 class, it is not, as it diverges at the transition point and
in the depinned phase.  Therefore equation (\ref{mn2}) {\it is not an
order-parameter Langevin equation} and it is $m=1/n$ that is to be
monitored once the equation for $n$ is integrated. Typically, Langevin
equations are written in terms of a vanishing-at-the-transition
order-parameter so that series expansions, truncation of power-series
to lowest orders can be employed and the applicability of standard
perturbative techniques is viable. This is in contrast with
equation (\ref{mn2}). Furthermore, the noise amplitude diverges at the
transition point.  This apparent ill-behavior may be the reason why
equation (\ref{mn2}) has been ignored so far in the literature.

In what follows we study the non-order parameter Langevin equation for
the MN2 class, equation (\ref{mn2}), using standard mean-field
approaches and integrating it numerically. Despite of the presence of
apparent pathologies and divergences at (and above) the transition
point, we find that equation (\ref{mn2}) reproduces the previously
known results for this universality class. In passing, we improve the
numerical precision of the corresponding critical exponents. This
constitutes a step forward in the general understanding of
nonequilibrium phase transitions into absorbing states, allows for a
better comparison with the MN1 class, gives a new starting point for
future analytical approaches, and serves as an example of a phase
transition best characterized in terms of an equation for a diverging
field that is not an order-parameter.
 
\section{Mean-Field analysis}

Standard mean-field approaches usually neglect spatial and
noise-induced fluctuations. For Langevin equations characterizing
spatially extended systems with multiplicative noise, however, this
approximation has proved to be too crude and unreliable in any
dimension \cite{MN4} as both spatial structure and noise are relevant
features.  Therefore, more elaborated methods are required to obtain a
sound qualitative picture of the transition \cite{MN5,Birner,vdb}. A
mean-field approach tailored to account properly for the noise term
and the spatially-varying order-parameter consists in discretizing the
Laplacian term as $1/2d \sum_j(n_j-n_i)$, where $d$ is the system
dimensionality and the sum is over the nearest-neighbors of site $i$
\cite{Birner,vdb}. When these latter are substituted by their average
value $\n $, a closed Fokker-Planck equation (which involves no
approximation in the noise) is obtained for $P(n,\n)$. The stationary
solution of such an equation is found by imposing the self-consistency
requirement
\cite{vdb}
\begin{equation}
\n = {\int_0^\infty dn \ n \ P(n,\n) 
\over \int_0^\infty dn  \ P(n,\n)}.
\label{self-consitstency}
\end{equation}
Note that this approach preserves the crucial role played by the
multiplicative noise and includes the spatial coupling even if in an
approximate (self-consistent) way. A detailed discussion of the
results obtained with this method for the MN1 class can be found in
\cite{MN5}. Applying this procedure to equation (\ref{mn2}) one readily obtains
\begin{equation}
P(n,\n) \propto n^{
%\displaystyle{ 
{2 (a-D) \over \sigma^2}-1} \exp \bigg\{
-{2 b \over \sigma^2 q n^q}
-{2 D\n \over \sigma ^2 n}
\bigg\},
\end{equation}
and after defining $I_p(\n) \equiv \int dn \ n^p \ P(n,\n)$,
\begin{equation}
\n  = {I_1(\n ) \over I_0(\n )}, \qquad
{\bar m} = {I_{-1}(\n) \over I_0(\n )}.
\end{equation}
To evaluate the scaling behavior of $I_p$ near the transition, when
$\n$ becomes large, we first do the substitution $z=2 D\n/\sigma^2 n$
and then expand the newly generated term $\exp \{ -2 b/ (
\sigma^{2(1-q)} q) \times (z/2 D\n)^q\}$ to first order to obtain
\begin{equation} 
I_p(\n) \approx A_p \n^{p+\gamma}+B_p\n^{p+\gamma-q},
\end{equation}
where $\gamma=2 (a-D)/\sigma^2$ and $A_p$, $B_p$ are expressed using
the Gamma function as
\begin{eqnarray}
A_p &=& {\left( 2 D\over \sigma^2 \right)}^{\gamma+p}
\Gamma(-p-\gamma), \nonumber \\ B_p &=& -{\left( 2 D\over \sigma^2
\right)}^{p+\gamma-q} {2 b \over \sigma^{2} q} \Gamma(-p-\gamma+q).
\end{eqnarray}
A direct calculation then yields $\n \sim {|a+\sigma^2 / 2|}^{-1/q}$
and ${\bar m} \sim |a+\sigma^2 / 2|^{1/q}$.  Therefore, the
order-parameter critical exponent is $\beta=1/q$ \cite{long}.

It is instructive to compare these results with those obtained using
the same technique for equation (\ref{mn1}), $\beta=\max \left(
1/q,\sigma^2/(2 D)\right)$
\cite{Birner,MN5}. That is, in the MN1 case two possible values of
$\beta$, reminiscent of the strong- and weak-coupling regimes of the
KPZ dynamics \cite{HZ}, appear already at this self-consistent
mean-field level.  In contrast, for equation (\ref{mn2}), there is no
strong-coupling regime, which would be characterized by a
non-universal, noise-amplitude-dependent, value of $\beta$. This fact
is related to the presence of a single cut-off in the stationary
probability distribution, equation (\ref{mn2}), while two
different cut-offs, and therefore two different mechanisms controlling
the scaling, appear for equation (\ref{mn1}) \cite{MN5}. The
implications of this property, as well as its connections with the
high-dimensional behavior of the KPZ dynamics, will be analyzed elsewhere.

\section{Numerical integration of stochastic differential equations}

In order to integrate equation (\ref{mn2}) as efficiently as possible
we have employed a recently introduced {\it split-step} scheme for the
integration of Langevin equations with non-additive noise. In this
scheme, the Langevin equation under consideration is studied on a
lattice and separated in two parts: the first one includes only
deterministic terms and is integrated at each time-step using any
standard integration scheme: Euler, Runge Kutta, etc \cite{Maxi} (here
we have chosen a simple Euler algorithm). The output of this step is
used as the input to integrate (along the same time-step) the second
part which may include the linear deterministic term and, more
importantly, the noise. This is done by sampling in an exact way the
probability distribution function associated with this part of the
equation.  In the case under study (noise proportional to the field)
the second step corresponds to sampling a log-normal distribution
solution of $\partial_t n= a n + \sigma n \eta$
%$\partial_t n(r,t)= a n(r,t) + \sigma n(r,t) \eta(r,t)$ 
(for more details see \cite{lognorm} and
\cite{DCM}).
 Therefore the two-step algorithm is finally specified by:
\begin{equation}
n_1 (i,t)=n(i,t)+\left(b~n(i,t)^{1-q}+D
[n(i+1,t)+n(i-1,t)-2n(i,t)]
\right)dt
\end{equation}
%where $\nabla^{2} n(i,t)=n(i+1,t)+n(i-1,t)-2n(i,t)$ 
and,
\begin{equation}
n(i,t+dt)=n_1(i,t)~ \exp\left(a dt + \sigma~ \sqrt{dt}~ \eta \right)
\end{equation}
where $\eta$ is a random variable extracted from a Normal distribution
with zero-mean and unit variance. Note that the linear deterministic
term can be included in either the first or the second step, or
partially incorporated in both of them.

We have considered one-dimensional lattices, and fixed $\sigma=1$,
$b=1$, $D=0.1$, space-mesh $dx=1$, and time-mesh $dt =0.1$ (note that
in this scheme $dt$ can be taken larger than in usual integration
algorithms \cite{DCM}). As initial condition we take $n(r, t=0)=3$. We
take $q=4$ for all simulations except for results presented in figure
\ref{fig-2}(a) where we show that asymptotic results do not depend on
the value of $q$, as long as $q>0$.  We then iterate the dynamics by
employing the previous two-step integration algorithm, using parallel
updating, at each site $i$.

\begin{figure}[h!]
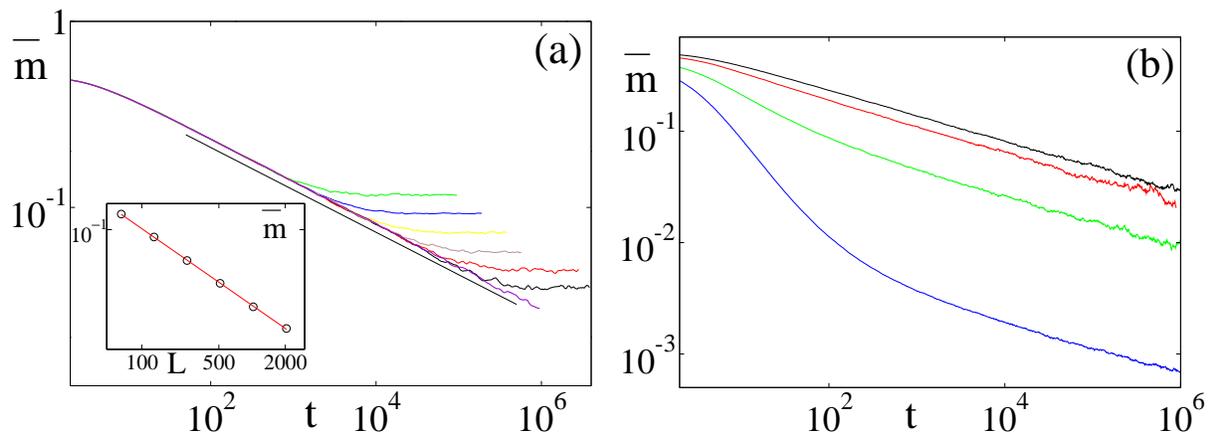

\centerline{
\epsfig{file=figure1a.eps,angle=0,width=8cm}
\epsfig{file=figure1b.eps,angle=0,width=8cm}}
\caption{A: Log-log plot of the time decay of the average order-parameter 
at the critical point, $a_c=-0.143668$, for system sizes $2^6$, $2^7$,
$2^8$, $2^9$, $2^{10}$, $2^{11}$, and $2^{17}$ respectively and $q=3$.
In the inset, the average saturation values of the previous curves are
plotted as a function of the system-size, $L$, in double-logarithmic
scale. B: Averaged order-parameter decay for equation (\ref{mn2}) with
$q=0.5, 1, 2, 4$ (from bottom to top) in a system of size
$L=2^{17}$. For any $q>0$ we observe the same asymptotic decay
exponent at the same critical point location.}
\label{fig-1}
\end{figure}

\begin{figure}[h!]
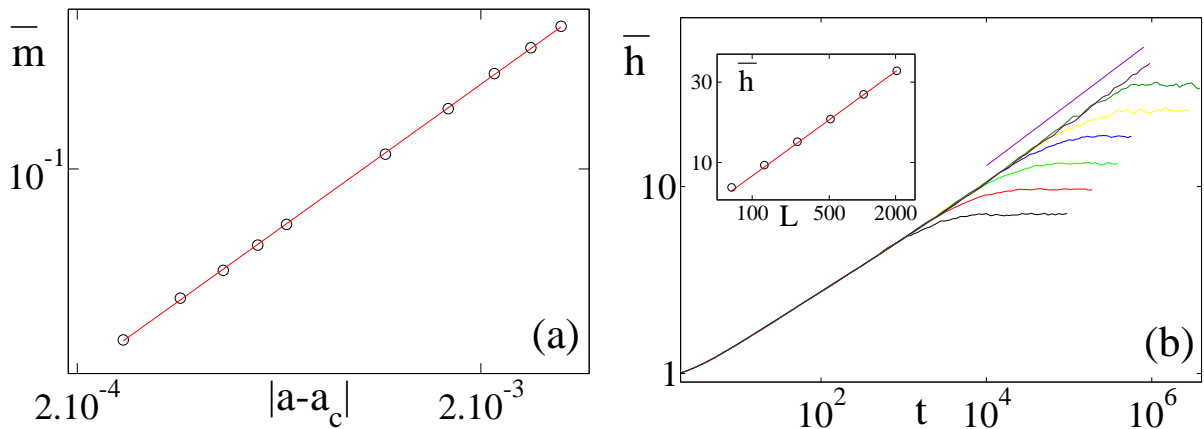

\centerline{
\epsfig{file=figure2a.eps,angle=0,width=8cm}
%\vspace{4.8cm}
\epsfig{file=figure2b.eps,angle=0,width=8cm}}
\caption{ 
A: Log-log plot of the saturation values of ${\bar m}$ for different
values of $a$ nearby the critical point.  From the slope we estimate
$\beta = 0.332(5)$.  B: Log-log plot of the time growth of the
averaged height $h=-\log(n)$ for the same value of $a$ and same system
sizes as in figure 1. In the inset, the average saturation values of
the previous curves are plotted as a function of the system-size, $L$,
in a log-log scale.}
\label{fig-2}
\end{figure}

A summary of our main findings follows. First, to accurately determine
the critical point, we perform decay experiments and average over many
independent runs in a system of size $L=2^{17}$.  At criticality,
$a_c=-0.143668(3)$, the average density, ${\bar m}=\overline{(1/n)}$
decays as a power-law with an associated exponent $\theta \approx
0.229(5)$ (see Fig.\ref{fig-1}). This is to be compared with the
previous estimates $\theta \approx 0.215(15)$
\cite{Fattah} and $\theta \approx 0.228(5)$ \cite{FSS-MN}.
On the other hand, for smaller system-sizes, we observe saturation at
this value of $a_c$, and the scaling of the saturation values for
different system sizes (inset Fig.\ref{fig-1}(a)) gives $\beta/\nu
\approx 0.335(5)$ (to be compared with $0.34(2)$ in \cite{Fattah}).

For other values of $q$ we have verified that, as shown in figure 2,
none of the previously reported exponents, nor the location of the
critical point, are altered, although $q-$dependent transient effects
exist. The invariance of $a_c$ against changes in $q$ can be
understood from the fact that $a_c$ corresponds to the value of $a$
for which depinned interfaces, arbitrarily far from the wall, invert
their direction of motion; this is not affected by the nature of the
wall, i.e. by the value of $q$.
Additionally, the order-parameter exponent $\beta \approx 0.332(5)$
has been measured for $L=2^{17}$ (see also figure 2). The previous
best estimation was $\beta=0.32(2)$ \cite{Fattah}.

We have also studied the scaling properties of the height field ${\bar
h} = - \overline{ \log (n) }$ (see Fig.\ref{fig-2}(b)).  An analysis
analogous to that presented above for $m$ leads to ${\bar h}(t)
\approx t^{0.323(10)}$, ${\bar h} \sim |a-a_c|^{-0.48(3)}$, and ${\bar
h}(a_c,L) \sim L^{0.48(2)}$, which define a set of critical exponents
for $h$ analogous to those for $m$ : $\theta_h \approx -0.323(10) $,
$\beta_h \approx -0.48(3)$, and $\beta_h/\nu \approx -0.48(2)$ (see
figure 2b). The previous best estimations for these exponents were:
$-0.355(15)$, $-0.52(2)$ and, $-0.52(2)$ respectively
\cite{Fattah}. The minus signs just reflect the fact that ${\bar h}$
diverges at the transition point.  A summary including our best
estimates for the critical exponents can be found in table 1.
\begin{table}[h!]
\centerline{
\begin{tabular}{|c|c|c|c|c|c|}
\hline
 & $\theta$ & $\beta$ & $\beta / \nu$ & $\nu$ & $z$  \\
\hline
 m & 0.229(5) & 0.332(5) & 0.335(5) & 0.99(3) &  1.46(5) \\
\hline
 h & -0.323(10) &  -0.48(3) & -0.48(2) & 1.0(1) & 1.48(10)  \\
\hline
% n \cite{Fattah} & 0.215(15) &  0.32(2) & 0.34(2) & 0.97(5) & 1.55(5)  \\
%\hline
% h \cite{Fattah} & 0.355(15) & 0.52(2) & 0.52 & 1 & 1.5  \\
%\cite{Fattah}
%\hline
\end{tabular}
}
\caption{Critical exponents for the MN2 class in $d=1$ as calculated
in this paper. Scaling of saturation times gives $z$ which is consistent
with the value $z=\beta / (\theta \nu)$ \cite{Review,MN1}. }
\label{tab1}
\end{table}
Finally, we stress that our results are compatible with the
theoretical predictions $z=3/2$ and $\nu=1$, expected to be valid for
both MN1 and MN2 \cite{MN1,Review}.  Also, all the standard scaling
laws between exponents are satisfied \cite{Review}.

\section{Discussion}

We have studied the dynamics of KPZ-like interfaces with $\lambda>0$
bounded by a lower-wall. The results and phenomenology differ from
that of upper-walls.  By performing a Cole-Hopf or logarithmic
transformation, the resulting order-parameter Langevin equation
(\ref{mn2bis}) is singular and no satisfying result can be derived
from it. Instead, the main result of this paper, is that a sound
Langevin equation (\ref{mn2}) can be written in terms of a
non-order-parameter field which diverges at the transition. For such
an equation we have performed (i) a self-consistent mean-field
analysis leading to the result $\beta=1/q$, and no trace of any
strong-coupling regime (noise-dependent $\beta$ exponent value)
contrarily to what happens for the upper-wall case; (ii) extensive
numerical integrations of the stochastic equation using a recently
introduced very-efficient numerical scheme. The obtained critical
exponent values are in good agreement with previously known ones
measured in simulations of discrete models, and improve the level of
accuracy and precision.

In summary, we have shown that an apparently ill-behaved
non-order-parameter Langevin equation constitutes a sound continuous
representation of the pinning-depinning transition experienced by
interfaces in the Kardar-Parisi-Zhang class under the presence of a
bounding lower-wall. Performing further analytical, renormalization
group analyses of the present Langevin equation remains as a
challenging task.

\ack%{\bf Acknowledgments}
%\vspace{0.1cm}

Financial support from MEyC and FEDER (project FIS2005-00791) is
acknowledged.

\section*{References}
%\vspace{0.1cm}


\begin{thebibliography}{50}


\bibitem{KPZ} 
Kardar M, Parisi G and Zhang Y C 1986 {\it Phys. Rev. Lett.} {\bf 56} 889

\bibitem{HZ}
Halpin-Healy T and Zhang Y C 1995 {\it Phys. Rep.} {\bf 254} 215
\nonum 
Barab\'{a}si A L and Stanley H E 1995 {\it Fractal Concepts in Surface
Growth} (Cambridge: University Press Cambridge) and references therein.

\bibitem{Review}
Mu\~noz M A 2004 {\it Nonequilibrium Phase Transitions and
Multiplicative Noise} (Advances in Condensed Matter and Statistical
Mechanic) ed E Korutcheva and R Cuerno (New York: Nova Science
Publishers) p~34 ({\it Preprint} cond-mat/0303650)

\bibitem{MN1} 
Grinstein G, Mu\~noz M A and Tu Y 1996 {\it Phys. Rev. Lett.} {\bf 76} 4376

\bibitem{MN2}
Tu Y, Grinstein G and Mu\~noz M A 1997 {\it Phys. Rev. Lett.} {\bf 78} 274

\bibitem{MN3}
Mu\~noz M A and Hwa T 1998 {\it Europhys. Lett.} {\bf 41} 147

\bibitem{Fattah} 
Mu\~noz M A, de los Santos F and Achahbar A 2003 {\it Braz. J. Phys.} {\bf 33} 443 ({\it Preprint} cond-mat/0304239)
%{\it Critical behavior of a type of bounded, Kardar-Parisi-Zhang equation},


\bibitem{Ahlers} 
Ahlers V and Pikovsky A 2002 {\it Phys. Rev. Lett.} {\bf 88} 254101
\nonum
Droz M and Lipowski A 2003 {\it Phys.  Rev. E} {\bf 67} 056204
%From multiplicative noise to directed percolation in wetting transitions
\nonum
Ginelli F, Ahlers V, Livi R, Mukamel D, Pikovsky A, Politi A, and Torcini A 2003 {\it Phys. Rev. E} {\bf 68} 065102
% V. Ahlers, R. Livi, D. Mukamel, A. Pikovsky, A. Politi, and A. Torcini


\bibitem{Romu} 
Mu{\~n}oz M A and Pastor Satorras R 2003 {\it Phys. Rev. Lett.} {\bf 90} 204101
%{\it Stochastic theory of synchronization transitions},

\bibitem{Lisboa}
de los Santos F, Telo da Gama M M and Mu\~noz M A 2002 {\it Europhys. Lett.} {\bf 57} 803
\nonum
de los Santos F, Telo da Gama M M and Mu\~noz M A 2003 {\it Phys. Rev. E} {\bf 67} 021607


\bibitem{Haye1}
Hinrichsen H, Livi R, Mukamel D and Politi A 1997 {\it Phys. Rev. Lett.} {\bf 79} 2710
%Wetting under nonequilibrium conditions
\nonum
Hinrichsen H, Livi R, Mukamel D and Politi A 2003 {\it Phys. Rev. E} {\bf 68} 041606
%Non-equilibrium  wetting of sinite samples,
\bibitem{FSS-MN}
Kissinger T, Kotowitz A, Kurz O, Ginelli F, and Hinrichsen H 2005 {\it J. Stat. Mech.} P06002 ({\it Preprint} cond-mat/0503582)

\bibitem{nota} %12
Indeed, it can be easily shown that a KPZ equation with positive
non-linearity and a ``lower-wall'' is equivalent (can be mapped by
changing the sign of $h$) to a KPZ with a negative non-linearity
coefficient and an ``upper-wall'' \cite{Review}.

\bibitem{nota2} In the more general case $\lambda \neq D$, the 
different coefficient can be reabsorbed by using 
$n=\exp({\lambda \over D} h)$.


\bibitem{Ito}%13
 The sole difference between utilizing the Ito or the 
Stratonovich calculus, in this case, is a trivial shift in $a$
\cite{VK}.

\bibitem{VK} %14
N G van Kampen 1981 {\it Stochastic Processes in Physics and Chemistry} (Amsterdam: North-Holland)
\nonum
%\bibitem{Gardiner} 
C W Gardiner 1985 {\it Handbook of Stochastic Methods} (Berlin and Heidelberg: Springer Verlag)

\bibitem{MN4} %15
Genovese W and Mu\~noz M A 1999 {\it Phys. Rev. E} {\bf 60} 69

\bibitem{Haye-MF} %16
Ginelli F and Hinrichsen H 2004 {\it J. Phys. A} {\bf 37} 11085


\bibitem{factor2} %16 
In fact, except for the factor $2$ in front of
$(\nabla m)^2 /m$ equation (\ref{mn2bis}) coincides with the Cole-Hopf
transform of $\partial_t h (x,t) = \nabla^2 h + a +b e^{-ph} +
\eta(x,t)$ that is the growth of wetting layers toward their
equilibrium state [Lipowsky R 1985 {\it J. Phys. A} {\bf 18}
L585]. Observe that this is just the equilibrium, Edwards-Wilkinson
model, in the presence of a bounding wall. Note also that the factor
$2$ in equation (\ref{mn2bis}) cannot be readsorbed by
reparametrizing.



\bibitem{KPZproblem} % 17
Newman T J and Bray A J 1996 {\it J. Phys. A} {\bf 29} 7917
\nonum
Lam C H and Shin F G 1998 {\it Phys. Rev. E} {\bf 58} 5592

\bibitem{MN5} % 18 
Mu\~noz M A, Colaiori F and Castellano C 2005 Mean field approach to
systems with multiplicative noise {\it Preprint} cond-mat/0506635. To
appear in {\it Phys. Rev. E}

\bibitem{Birner} %19
Birner T, Lippert K, M{\"u}ller R, K{\"u}hnel A, and Behn U 2002 {\it Phys. Rev. E} {\bf 65} 046110

\bibitem{vdb} % 20
Van den Broeck C, Parrondo J M R, Armero J and Hern\'andez Machado A 1994 {\it Phys. Rev. E} {\bf 49} 2639

\bibitem{long} % 21 
For some values of $q$, $I_p(\n)$ can be evaluated
exactly: $q=1$ immediately leads to $\n=-2 b/(a+{\sigma^2/2})\sim
\epsilon^{-1}$ and ${\bar m} = \gamma (a+\sigma^2/2)/b(1+\gamma) \sim
\epsilon$; for $q=2$, $I_p$ can be expressed in terms of
parabolic-cylinder functions $D_\nu(x)$. The asymptotic expansion for
large $x$ $D_\nu(x) \approx e^{-x^2/4}x^\nu(1-\nu(\nu+1)/x^2+...)$
leads to $\n \sim \epsilon ^{-1/2}$ and ${\bar m \sim
\epsilon^{1/2}}$. 
%We have also numerically solved equation(\ref{self-consitstency})
%and checked that indeed $\beta=1/q$ for $q=0.5,1,2,10$.

\bibitem{Maxi} % 22
San Miguel M and Toral R 1997 {\it Stochastic Effects in
Physical Systems} (Instabilities and Nonequilibrium
Structures VI) ed Tirapegui E and Zeller W (Kluwer Academic
Publishers) pp~35-130 ({\it Preprint} cond-mat/9707147)


\bibitem{lognorm} % 23
The part of the Langevin equation to be integrated can be written as $
dn_t=a n_t{\it dt}+\sigma n_t{\it dW_t}$ where $dW$ is a Wiener
process. Since this is interpreted in the Stratonovich sense, we can
safely perform the change of variables $Y_t=\ln n_t$ and obtain $
dY_t=a{\it dt}+\sigma{\it dW_t}$.  This describes a drifted Brownian
motion equation whose solution is given by a normal distribution of
mean $y_0+adt$ and variance $\sigma^{2}dt$:
$Prob(Y_{t+dt}=y|Y_t=y_0)=N(y_0+adt,\sigma^{2}dt)$. Inverting the
change of variables, we are left with a log-normal form, which can be
sampled in an exact way by taking:
%
%\begin{equation}
%  Proba(X_{t+dt}=y|X_t=X_0)= 
% Proba(Y_{t+dt}=y=ln(x)|Y_t=y_0=ln(x_0))|\frac{dY}{dX}|
%\end{equation}
%and changing  back variables:
%\begin{equation}
%  Proba(X_{t+dt}=y|X_t=X_0)= X_0exp(adt+\sigma\sqrt{dt}N(0,1)).
%\end{equation}
%
%This is exactly sampled in the following way
$ n[t+dt|n(t)=n_0]=n_0~\exp(adt+\sigma\sqrt{dt}\eta)$. This same
expression can also be derived by changing variables in the Langevin
equation, performing one time-step evolution, and changing back
variables.

\bibitem{DCM} % 24
Dornic I, Chat\'e H and Mu\~noz M A 2005 {\it Phys. Rev. Lett.} {\bf
94} 100601
%{\it Integration of Langevin Equations with Multiplicative Noise
%and Viability of Field Theories for Absorbing Phase Transitions},

\nonum
Dornic I, Chat\'e H and Mu\~noz M A {\it In preparation}
\nonum
Pechenik L and Levine H 1999 {\it Phys. Rev. E} {\bf 59} 3893
\nonum
Moro E 2004 {\it Phys. Rev. E} {\bf R 70} 045102

\end{thebibliography}
\end{document}